\documentclass{webofc}
\usepackage[varg]{txfonts}   

\begin{document}
\title{Latest ALICE results on charm and beauty hadronisation in hadronic collisions}
%
%

\author{\firstname{Jianhui} \lastname{Zhu}\inst{1,2}\fnsep\thanks{\email{jianhui.zhu@cern.ch}} for the ALICE Collaboration
}

\institute{Institute of Particle Physics, Central China Normal University, 152 Luoyu Road, 430079 Wuhan, China
\and
           Istituto Nazionale di Fisica Nucleare, Sezione di Padova, Via Marzolo 8, 35131 Padova, Italy
          }

\abstract{%
    The study of heavy-flavour mesons and baryons in hadronic collisions provides unique access to the properties of heavy-quark hadronisation in the presence of large partonic densities, where new mechanisms of hadron formation beyond in-vacuum fragmentation can emerge. Performing these measurements in intervals of charged-particle multiplicities across different collision systems provides sensitivity to understand whether different hadronisation mechanisms are at play in small and large hadronic colliding systems.
    In this contribution, a selection of the latest charm and beauty production measurements in proton--proton (pp) collisions is presented, which can shed light on the modification of the heavy-quark hadronisation mechanisms with respect to leptonic collisions. New published results of the production of all prompt charm ground states in pp collisions at $\sqrt{s}=13$~TeV, which allowed us to measure the charm fragmentation fractions and the total $\rm c\bar{c}$ production cross section at midrapidity, will be shown. The new final measurement of non-prompt (i.e. originating from beauty-hadron decays) $\rm \Lambda_c^+$ baryons in the same collisions system will be discussed to provide a quantitative comparison between the hadronisation properties of beauty and charm hadrons. New measurements of $\rm \Xi_c^0$ production as a function of charged-particle multiplicity in pp collisions at $\sqrt{s}=13$~TeV, and of $\rm \Xi_c^0$ production in p--Pb collisions at $\sqrt{s_{\rm NN}}=5.02$~TeV, will be also presented, shedding further light on the hadronisation of charm-strange baryons in different colliding systems.
}
\maketitle
\section{Introduction} \label{intro}
Measurements of heavy-flavour hadron production in hadronic collisions provide crucial tests for calculations based on quantum chromodynamics (QCD). Typically, the calculations of transverse momentum ($p_{\rm T}$) differential heavy-flavour hadron production cross sections in hadronic collisions are factorised into three independent components \cite{COLLINS198637}: the parton distribution functions (PDFs), which describe the Bjorken-$x$ distributions of the scattering partons within the incoming hadrons; the hard-scattering cross section for the partons to produce a charm or beauty quark, calculated as a perturbative expansion in powers of the strong coupling constant $\alpha_{\rm s}$, and the fragmentation functions, which characterise the hadronisation of a quark into a given hadron species. The latter term cannot be calculated with perturbative QCD (pQCD) methods, and is usually determined from measurements in $\rm e^+e^-$ collisions. It is then exploited in cross section calculations under the assumption that the relevant hadronisation processes are “universal”, i.e. independent of the collision energy and system. Hadron-to-hadron production ratios within the heavy-flavour sector, such as prompt $\rm \Lambda_c^+/D^0$ and non-prompt $\rm \Lambda_c^+/D^0$, are especially effective for probing hadronisation effects, since in theoretical calculations the PDFs and the partonic interaction cross sections are common to all heavy-flavour hadron species and their effects cancel in the yield ratios.

\section{Results} \label{sec-2}
\subsection{Prompt and non-prompt $\rm \Lambda_c^+/D^0$ yield ratio in pp and p--Pb collisions} \label{sec-2-1}
\begin{figure}[h]
\centering
\includegraphics[width=0.49\textwidth]{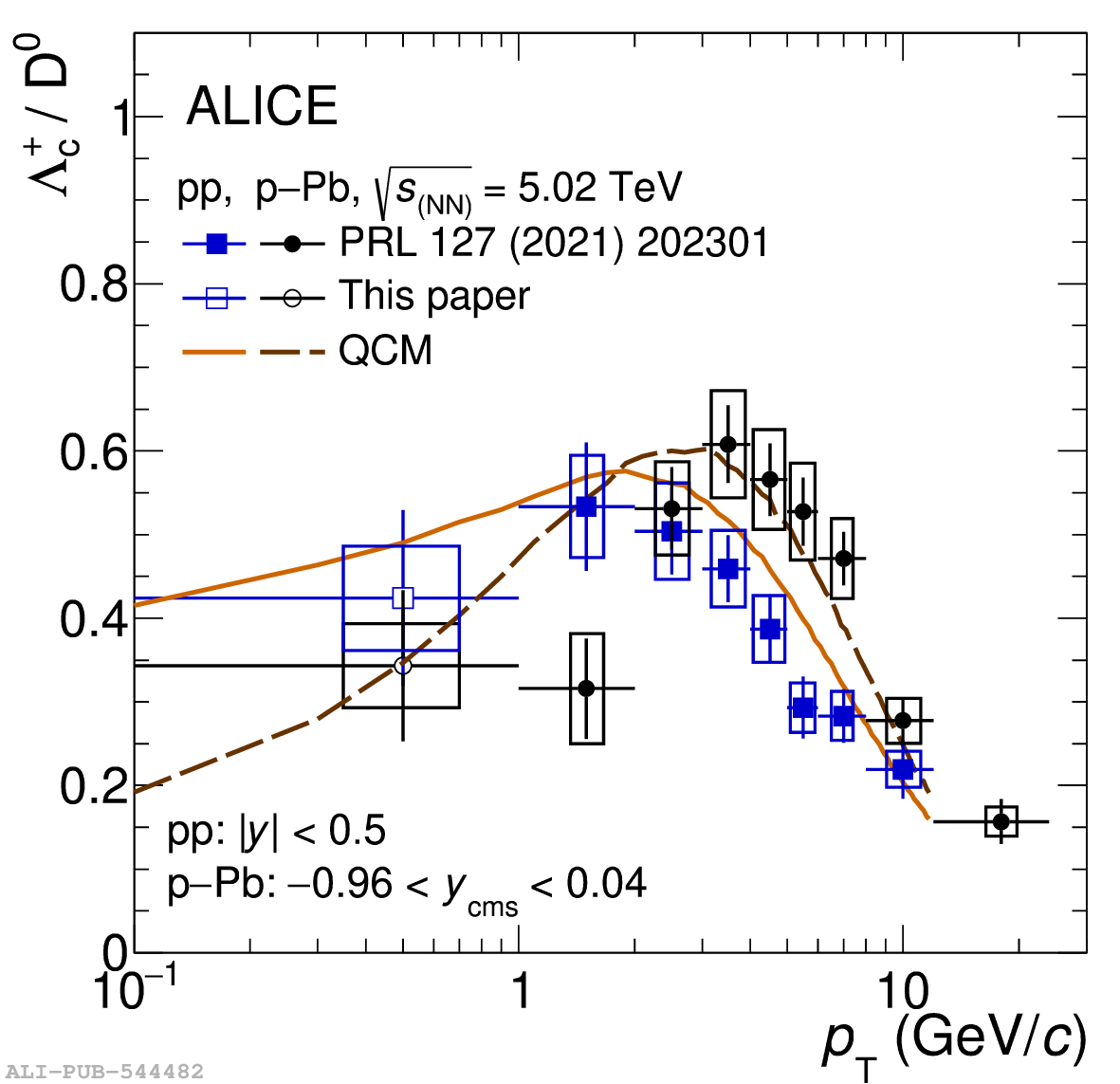}
\includegraphics[width=0.49\textwidth]{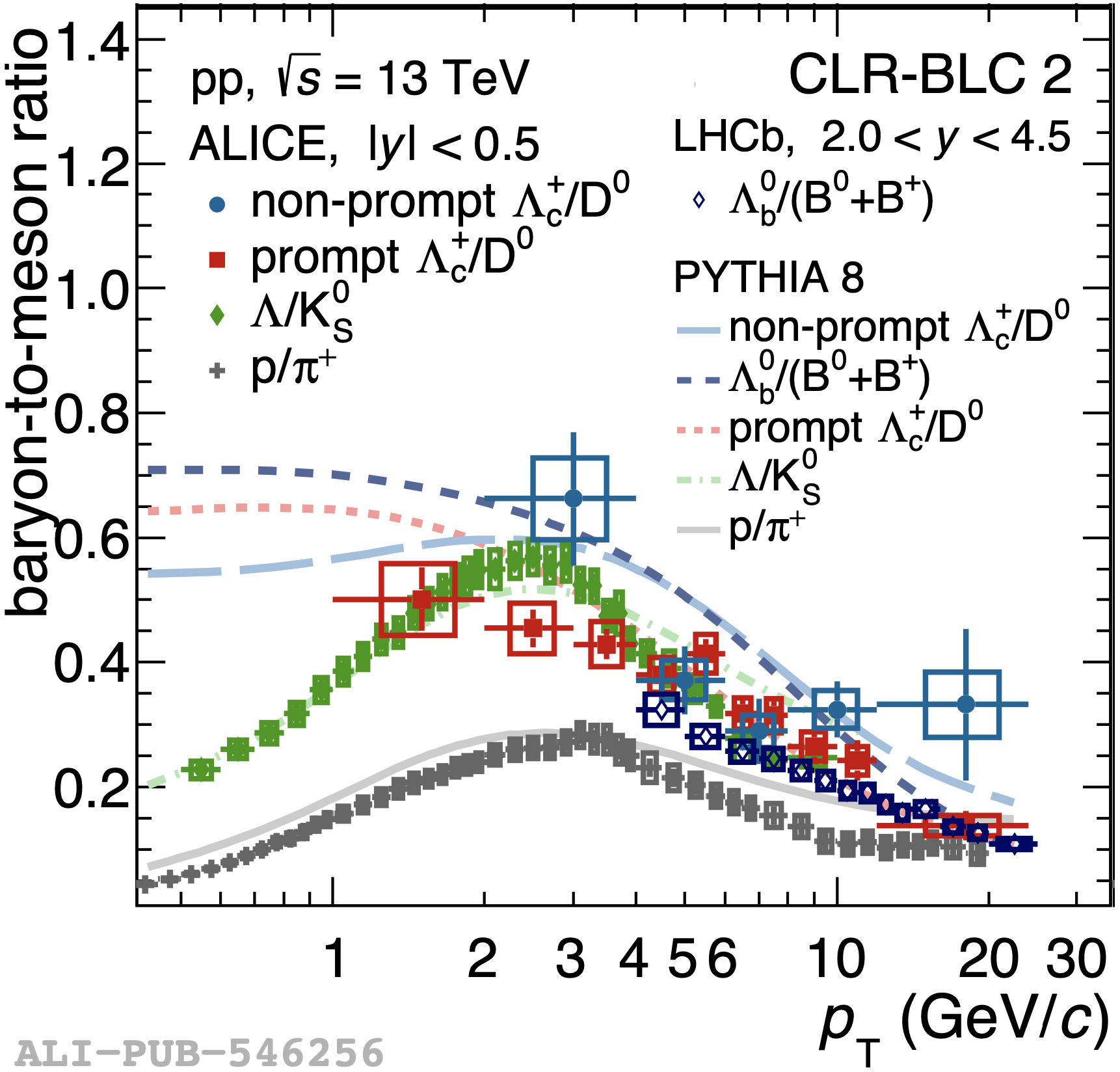}
\caption{Left: prompt $\rm \Lambda_c^+/D^0$ yield ratio in pp and p--Pb collisions as a function of $p_{\rm T}$ \cite{ALICE:2022exq}, compared with QCM expectations \cite{Li:2017zuj, Li:2021nhq}. Right: non-prompt $\rm \Lambda_c^+/D^0$ \cite{ALICE:2023wbx}, prompt $\rm \Lambda_c^+/D^0$ \cite{ALICE:2021rzj}, $\rm \Lambda/K_s^0$ \cite{ALICE:2020jsh}, and $\rm p/\pi^+$ \cite{ALICE:2020jsh} yield ratios measured in pp collisions at $\sqrt{s}=13$~TeV, at midrapidity ($|y|<0.5$), compared with the $\rm \Lambda_b^0/(B^0+B^+)$ yield ratio measured by LHCb at forward rapidity ($2.0<y<4.5$) \cite{LHCb:2019fns} and with predictions obtained with the PYTHIA 8 MC generator with color-reconnection mechanisms beyond the leading-color (CLR-BLC) approximation \cite{Christiansen:2015yqa} at midrapidity.}
\label{fig-1}
\end{figure}

The left panel of Fig.~\ref{fig-1} shows the prompt $\rm \Lambda_c^+/D^0$ production yield ratio measured at midrapidity as a function of $p_{\rm T}$ in pp and p--Pb collisions at $\sqrt{s_{\rm NN}}=5.02$~TeV down to $p_{\rm T}=0$. The shift of the peak towards higher $p_{\rm T}$ in p--Pb collisions could be attributed to a contribution of collective effects, e.g. radial flow. The results are compared with expectations from the QCM model \cite{Li:2017zuj, Li:2021nhq}, that includes a contribution of hadronisation of charm quarks via coalescence. The model describes the magnitude of the $\rm \Lambda_c^+/D^0$ yield ratio well for $p_{\rm T}<12$~GeV/$c$ in both collision systems and predicts the shift of the peak towards higher $p_{\rm T}$, resulting from a hardening of the $\rm \Lambda_c^+$ spectrum in p--Pb collisions. The $p_{\rm T}$-differential non-prompt $\rm \Lambda_c^+/D^0$ yield ratio at midrapidity in pp collisions at $\sqrt{s}=13$~TeV, compared with the measurements of prompt $\rm \Lambda_c^+/D^0$, $\rm \Lambda/K_s^0$, and $\rm p/\pi^+$ ratios at the same energy and rapidity interval and with the $\rm \Lambda_b^0/(B^0+B^+)$ yield ratio measured by LHCb at forward rapidity ($2.0<y<4.5$), is shown in the right panel of Fig.~\ref{fig-1}. The measured baryon-to-meson ratios of light flavour, strange, charm, and beauty hadrons show a similar $p_{\rm T}$ trend. The results are compared with PYTHIA 8 simulations with CR-BLC Mode 2 tune that include color-reconnection mechanism beyond leading-color approximation in the string fragmentation. Both prompt and non-prompt $\rm \Lambda_c^+/D^0$ yield ratio at low and intermediate $p_{\rm T}$ are substantially enhanced compared to measurements in $\rm e^+e^-$ or $\rm e^-p$ collisions.

\subsection{Prompt $\rm \Xi_c^0/D^0$ yield ratio in pp and p--Pb collisions} \label{sec-2-2}
\begin{figure}[h]
\centering
\includegraphics[width=0.44\textwidth]{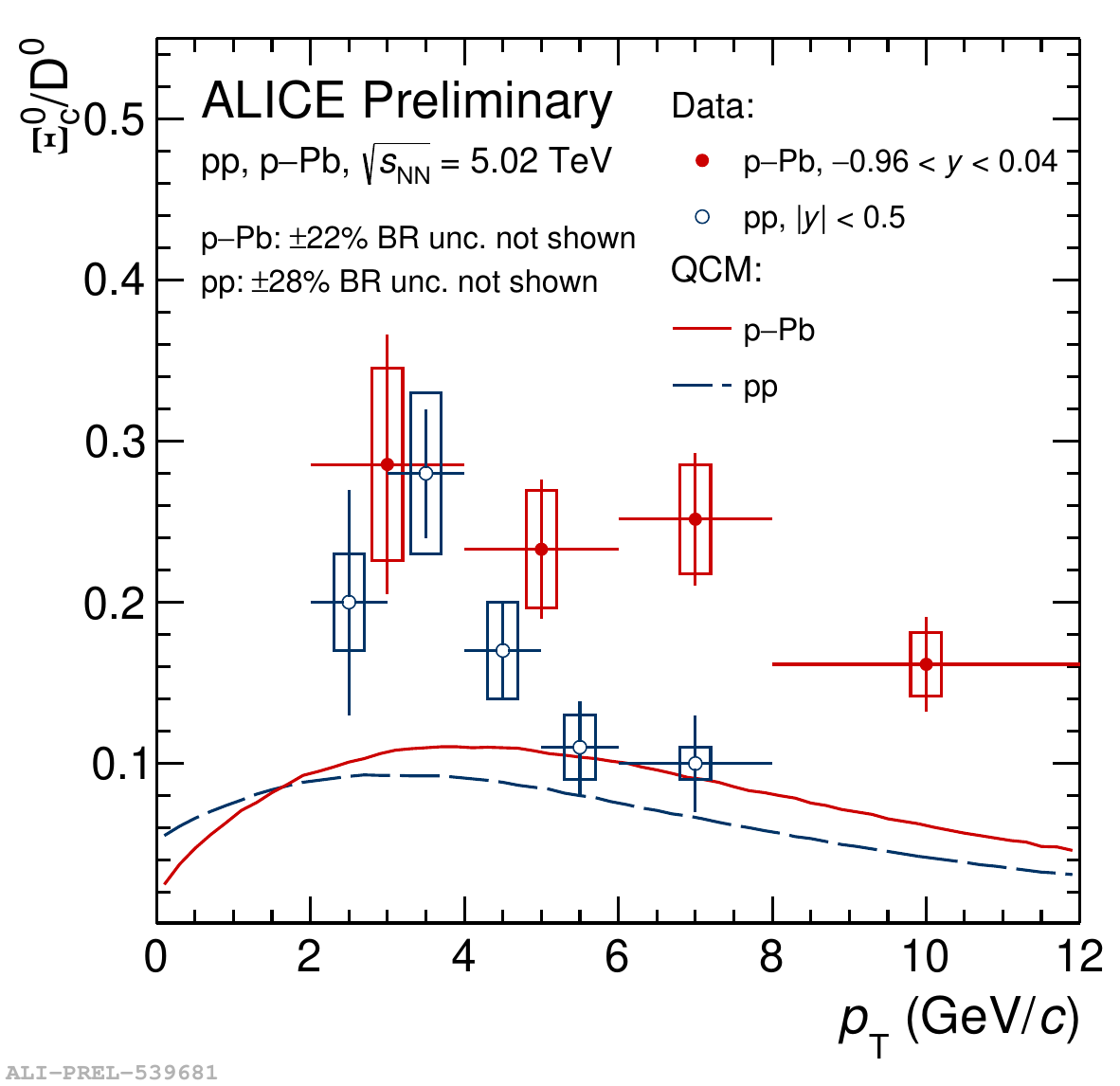}
\includegraphics[width=0.55\textwidth]{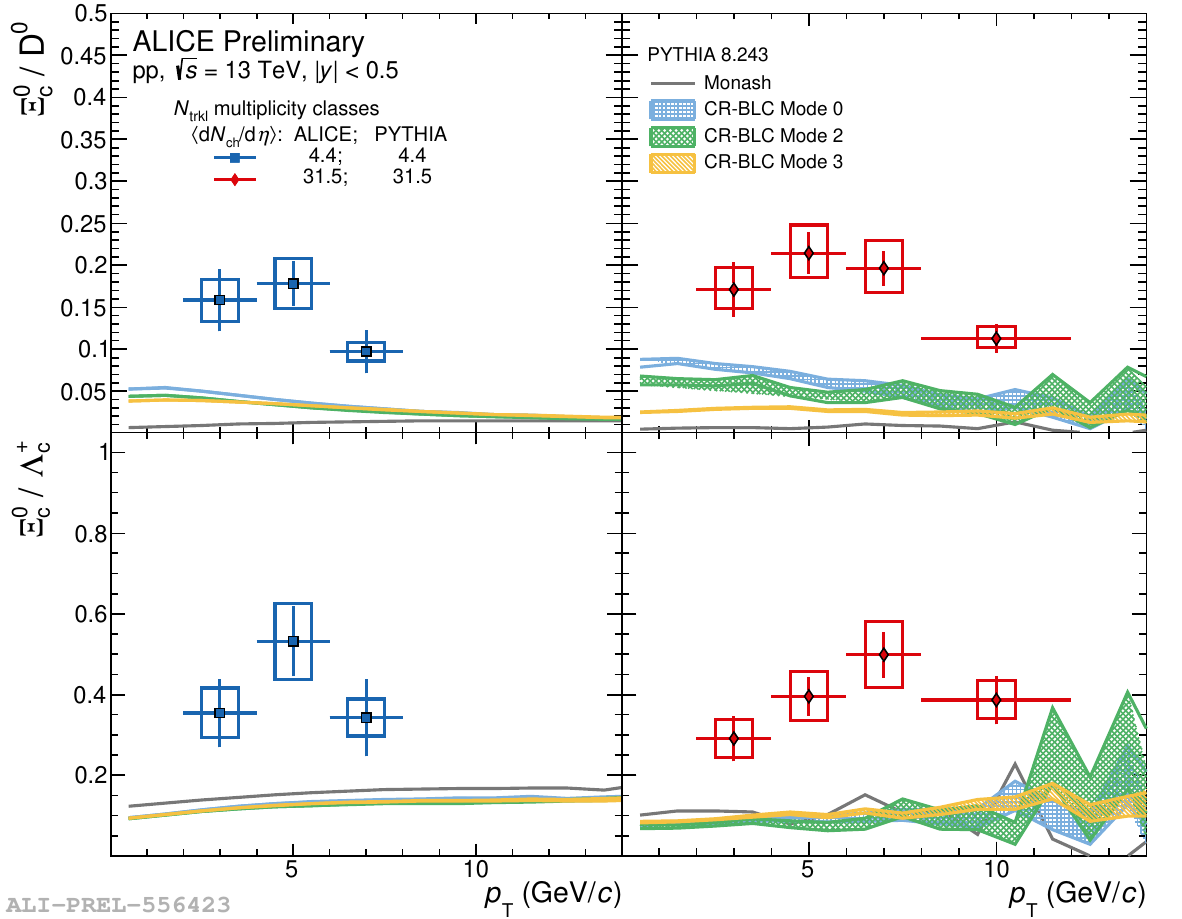}
\caption{Left: $\rm \Xi_c^0/D^0$ yield ratio as a function of $p_{\rm T}$ in pp and p--Pb collisions at $\sqrt{s_{\rm NN}}=5.02$~TeV compared to QCM predictions \cite{Li:2017zuj, Li:2021nhq}. Right: $\rm \Xi_c^0/D^0$ (top) and $\rm \Xi_c^0/\Lambda_c^+$ (bottom) yield ratios measured in pp collisions at $\sqrt{s}=13$~TeV at low (left) and high (right) multiplicity at midrapidity. The measurements are compared to PYTHIA predictions with Monash \cite{Skands:2014pea} and CR-BLC tunes \cite{Christiansen:2015yqa}, estimated in similar multiplicity classes.}
\label{fig-2}
\end{figure}

Similarly to the $\rm \Lambda_c^+/D^0$ yield ratio, the $\rm \Xi_c^0/D^0$ yield ratio was measured in pp and p--Pb collisions as a function of $p_{\rm T}$. The results are shown in the left panel of Fig.~\ref{fig-2} and compared with QCM expectations \cite{Li:2017zuj, Li:2021nhq}. The $\rm \Xi_c^0/D^0$ yield ratio in p--Pb collisions shows a slightly decreasing trend with $p_{\rm T}$, albeit with large uncertainties, and the $p_{\rm T}$ dependence is less pronounced than in pp collisions. The QCM predictions tend to underestimate the $\rm \Xi_c^0/D^0$ yield ratios in both collision systems. In the right panel of Fig.~\ref{fig-2}, the $\rm \Xi_c^0/D^0$ and $\rm \Xi_c^0/\Lambda_c^+$ yield ratios measured in pp collisions at $\sqrt{s}=13$~TeV in a low and a high multiplicity class at midrapidity are shown. They are underestimated by PYTHIA predictions with Monash \cite{Skands:2014pea} and CR-BLC tunes \cite{Christiansen:2015yqa}. From lowest to highest multiplicity class, there is no significant multiplicity dependence as a function of $p_{\rm T}$ within uncertainties for $\rm \Xi_c^0/D^0$ and $\rm \Xi_c^0/\Lambda_c^+$ yield ratios.

\subsection{Charm production and fragmentation in pp collisions} \label{sec-2-3}
\begin{figure}[h]
\centering
\includegraphics[width=0.55\textwidth]{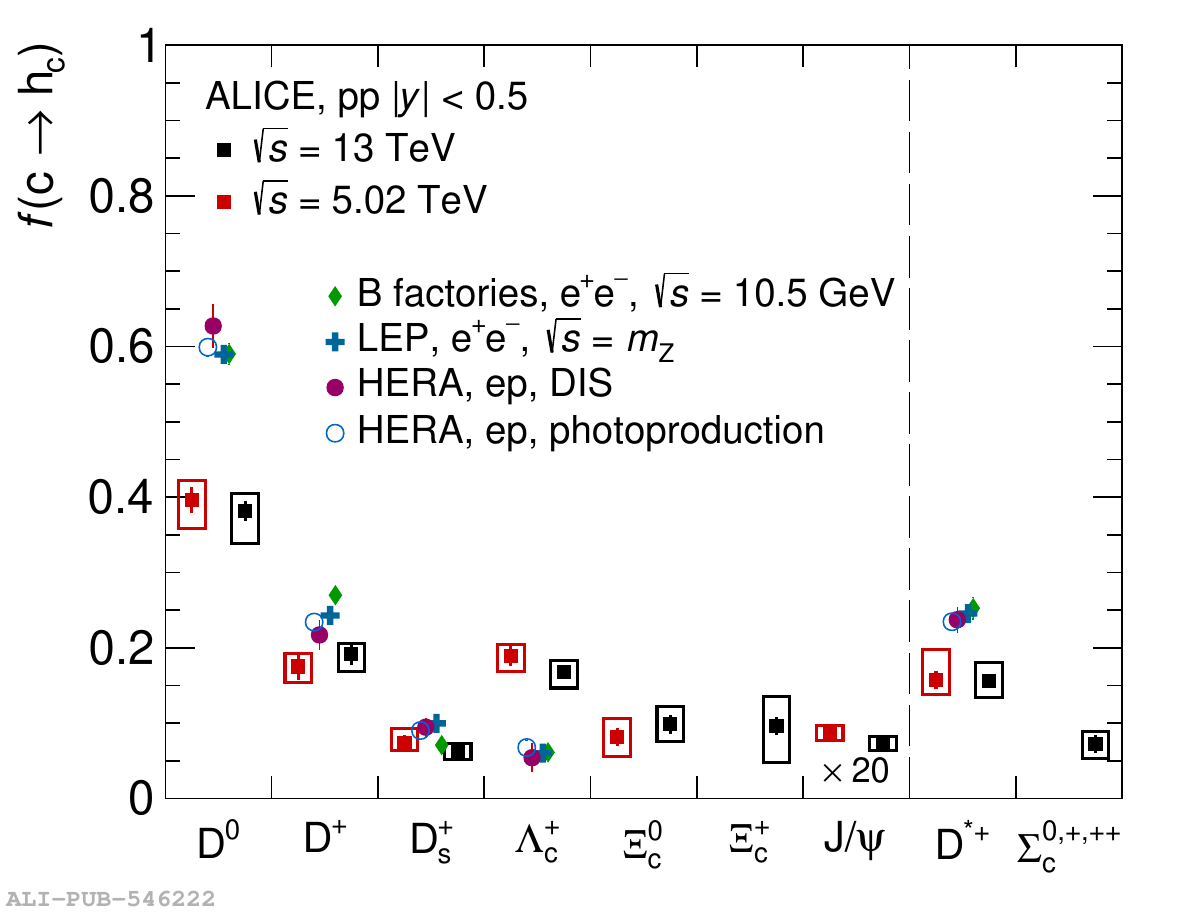}
\includegraphics[width=0.44\textwidth]{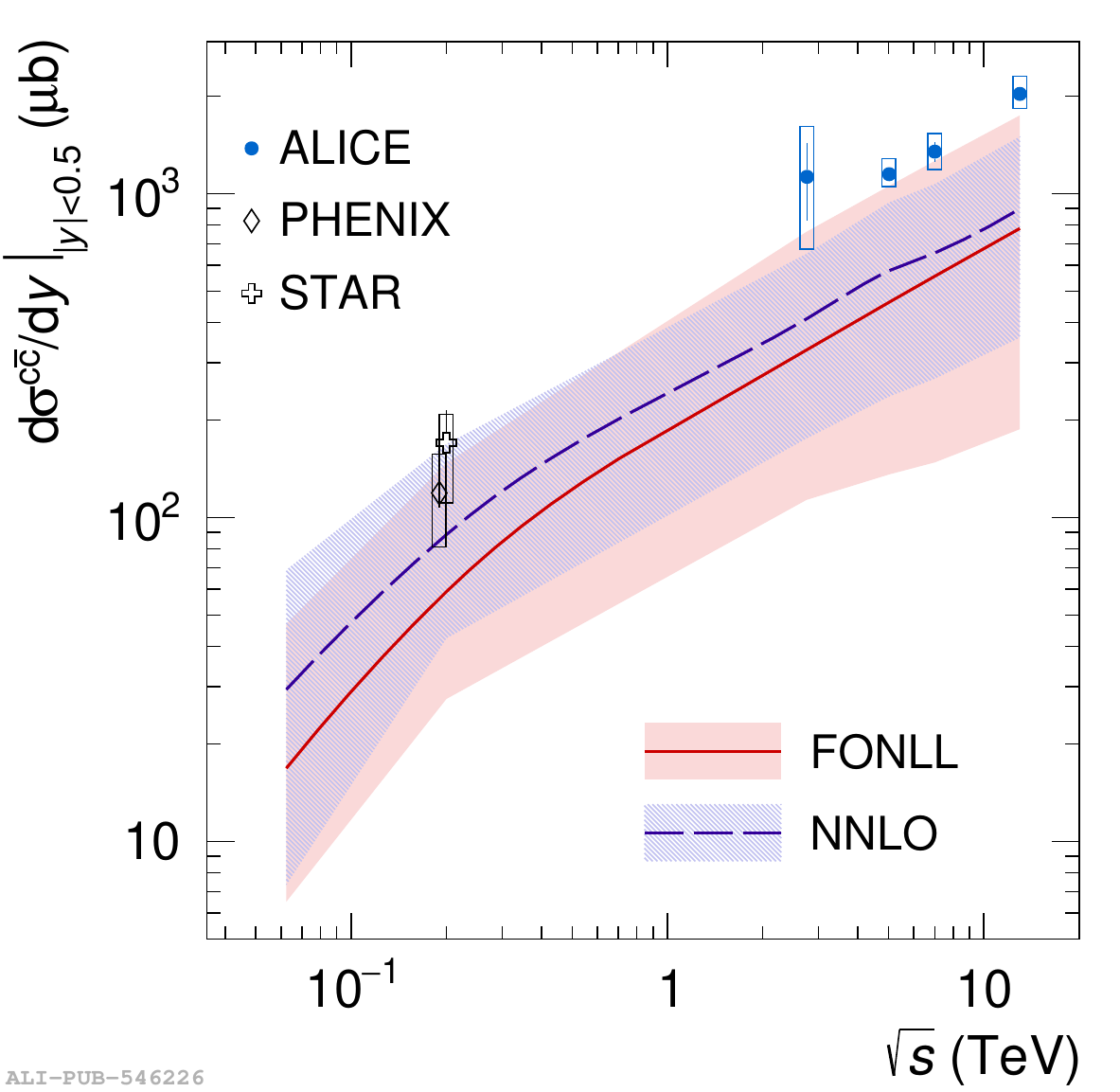}
\caption{Left: charm-quark fragmentation fractions into charm hadrons measured in pp collisions at $\sqrt{s}=5.02$~TeV and at $\sqrt{s}=13$~TeV, compared to experimental measurements performed in $\rm e^+e^-$ and ep collisions \cite{ALICE:2023sgl}. Right: total $\rm c\bar{c}$ production cross section at midrapidity per unit of rapidity as a function of the collision energy at the LHC \cite{ALICE:2023sgl} and RHIC \cite{STAR:2012nbd, PHENIX:2010xji}, compared to FONLL \cite{Cacciari:2012ny} and NNLO \cite{dEnterria:2016ids} calculations.}
\label{fig-3}
\end{figure}

The charm fragmentation fractions $f(\rm c\rightarrow H_c)$ shown in the left panel of Fig.~\ref{fig-3} for pp collisions at $\sqrt{s}=5.02$ and 13 TeV represent the probability of a charm quark to hadronise into a given charm-hadron species. The fraction of charm quarks that hadronise into baryons is about 40\%, around four times larger than what was measured at colliders with electron beams, showing that the assumption of the charm fragmentation universality is not valid. No significant energy dependence is observed for the fragmentation fractions within uncertainties.

The $\rm c\bar{c}$ production cross section per unit of rapidity at midrapidity ($\rm d{\it \sigma}^{c\bar{c}}/d{\it y}|_{|{\it y}| < 0.5}$) is calculated by summing the $p_{\rm T}$-integrated cross sections of all measured weakly-decaying charm hadrons.
The result is shown in the right panel of Fig.~\ref{fig-3} together with measurements at RHIC \cite{STAR:2012nbd, PHENIX:2010xji}. The measured cross sections lie on the upper edge of FONLL \cite{Cacciari:2012ny} and NNLO \cite{dEnterria:2016ids} predictions.

\section{Summary}
In this contribution, the latest ALICE results related to charm and beauty hadronisation mechanisms in hadronic collisions are reported. The assumption of universal parton-to-hadron fragmentation fractions is not valid at LHC energies. The heavy-quark hadronisation mechanisms in small collision systems at LHC need further investigation. More precise measurements of heavy-flavour baryon production with Run 3 data will provide stricter constraints on these mechanisms.

This work was supported by the National Natural Science Foundation of China (NSFC) (No. 12105109).

\let\oldthebibliography\thebibliography
\let\endoldthebibliography\endthebibliography
\renewenvironment{thebibliography}[1]{
  \begin{oldthebibliography}{#1}
    \setlength{\itemsep}{0.05em}
    \setlength{\parskip}{0em}
}
{
  \end{oldthebibliography}
}

\bibliography{references}

\end{document}